\documentclass{ws-ijmpd}

\usepackage{graphicx}  
\usepackage{latexsym}
\usepackage{cite}  

\def\beq{\begin{equation}}
\def\eeq{\end{equation}}

\def\rmd{{\rm d}}

\begin{document}

%

\def\nocropmarks{\vskip5pt\phantom{cropmarks}}

\let\trimmarks\nocropmarks      

%

\markboth{Bini D., Cherubini C., Geralico A. and Jantzen R. T.}
{Inertial effects in accelerating spacetimes
}

%
\catchline{}{}{}
%

\title{INERTIAL EFFECTS IN ACCELERATING SPACETIMES
}

\author{\footnotesize DONATO BINI}

\address{Istituto per le Applicazioni del Calcolo ``M. Picone'', CNR I-00161 Rome, Italy  \\
ICRA, University of Rome ``La Sapienza'', I-00185 Rome, Italy\\
INFN - Sezione di Firenze, Polo Scientifico, Via Sansone 1, I-50019, Sesto Fiorentino (FI), Italy 
\footnote{binid@icra.it}
}

\author{CHRISTIAN CHERUBINI}

\address{
Facolt\`a di Ingegneria, Universit\`a ``Campus Biomedico'', Via E. Longoni 47, I-00155 Rome, Italy\\
ICRA, University of Rome ``La Sapienza'', I-00185 Rome, Italy
\footnote{cherubini@icra.it}
}

\author{ANDREA GERALICO}

\address{
Physics Department and
ICRA, University of Rome ``La Sapienza,'' I-00185 Rome, Italy
\footnote{geralico@icra.it}
}

\author{ROBERT T. JANTZEN}

\address{
Department of Mathematical Sciences, Villanova University, Villanova, PA 19085, USA\\
ICRA, University of Rome ``La Sapienza'', I-00185 Rome, Italy
\footnote{robert.jantzen@villanova.edu}
}

\maketitle

\begin{history}
\received{
}
\revised{}
\end{history}

\begin{abstract}
The motion of test particles along circular orbits in the vacuum $C$ metric is studied
in the Frenet-Serret formalism. 
Special orbits and corresponding intrinsically defined geometrically relevant properties are selectively studied.
\end{abstract}


\section{Introduction}

General relativity asserts the equivalence between inertial and gravitational effects. However, our intuition often relies on Newtonian ideas. By re-introducing inertial forces through the consideration of relative motion, we can sometimes better interpret the behavior of geodesic or accelerated orbits in certain types of spacetimes. Most of the work done in introducing such a point of view has focused explicitly on transverse acceleration effects associated with centrifugal and Coriolis-like forces in the relativistic setting. The present investigation shows how these ideas are affected by uniform ``linear'' acceleration, for which the simplest nonflat exact solution of Einstein's vacuum equations
is the C metric. This metric represents the exterior field of a uniformly accelerated spherical gravitational source and it can also be thought of as a nonlinear superposition of the (flat) Rindler spacetime associated with a uniformly accelerated family of observers and the Schwarzschild solution representing a static black hole.  While the C metric spacetime is globally pathological and hence of limited value for physical interpretation, locally it can give some insight into the effects of \lq\lq acceleration of the spacetime" on transversely accelerated world lines similar to the study of the effects on such world lines of the \lq\lq rotation of the spacetime" due to the rotation of a Kerr black hole, for example.

Timelike geodesics were first discussed by Pravda and Pravdov\'a\cite{pravdacqg}  using various coordinate systems distinct from Boyer-Lindquist-like coordinates.
They constructed an effective potential, whose properties allowed them to distinguish three distinct types of timelike geodesics corresponding to particles which (a) fall into the black hole horizon, (b) cross the acceleration horizon and reach future infinity, or (c) spin around the $z$-axis, co-accelerating with the black hole and then reaching future infinity. 
They focused only on geodesics of the third type (circular geodesics) and numerically investigated the stability of timelike as well as null geodesics. 

In the present paper we use the Frenet-Serret formalism in Boyer-Linquist-like coordinates to study the motion of test particles along both geodesic and accelerated circular orbits and discuss the main geometrical and kinematical features of accelerated orbits with certain special geometrical properties.
Comparisons with the limiting cases of the Schwarzschild and Rindler spacetimes are carefully explored.

\section{Test particles along circular orbits in the C metric}

The vacuum C metric is a boost-rotation symmetric static spacetime of Petrov type D belonging to the Weyl class of solutions of the Einstein 
equations\cite{ES} and can be interpreted as the field of two black holes which are uniformly accelerated in opposite directions. Its original Kinnersley form\cite{kin69} is
\begin{equation}
\label{met_txyz}
\rmd s^2=\frac{1}{A^2(\tilde x+\tilde  y)^2}[(-\tilde F \rmd t^2 + \tilde F^{-1}\rmd \tilde y{}^2) +(\tilde G^{-1} \rmd \tilde x{}^2 + \tilde G\rmd \tilde z{}^2)]\ ,
\end{equation}
where
\begin{equation}
\tilde F(\tilde y)  = -1+\tilde y{}^2-2MA\tilde y{}^3,\quad  
\tilde G(\tilde  x) = 1-\tilde x{}^2-2MA\tilde x{}^3 = -\tilde F(-\tilde x)\ .
\end{equation}
The $(t,\tilde x,\tilde y, \tilde z)$ coordinates are adapted to the hypersurface-orthogonal Killing vector $\kappa=\partial_t$ and to the spacelike Killing vector
$\partial_{\tilde z}$. It turns out that $\partial_{\tilde x}$ is aligned with the non-degenerate eigenvector of the spatial Ricci tensor.

The constants $M\ge 0$ and $A\ge 0$ denote the mass and acceleration of the source, respectively, whose product $MA$ is dimensionless. 
The metric (\ref{cmetu}) is a nonlinear superposition of two metrics which are special cases, one  associated with the Schwarzschild black hole (the case $A=0$) and the other flat spacetime in uniformly accelerating coordinates (the case $M=0$).\cite{kin69,kinwal,Farh,hongteo}
Moreover, the C metric as written is assumed to have signature +2 with $\tilde F>0, \tilde G>0$, which limit the ranges of the nonignorable coordinates. To avoid a naked singularity, one must restrict the two parameters by the condition $MA<1/(3\sqrt{3})\approx 0.1925$.\cite{Farh,pavda,podol} For the sake of illustration, all of the figures below are drawn for the parameter value $MA=0.1$, which is roughly half this value.
Units are chosen such that the gravitational constant and the speed of light in vacuum are unity.

It is useful to introduce the retarded coordinate $u$, the radial coordinate $r$ and the azimuthal coordinate $\phi$
define by the relations\cite{kinwal,Farh}
\begin{equation}
u=\frac1A [t+\int^{\tilde y} \tilde F^{-1}\rmd \tilde y]\ , 
\qquad 
r=\frac{1}{A(\tilde  x+\tilde  y)}, \qquad \phi= \tilde z\ ,
\end{equation}
so that the metric takes the form
\begin{equation}
\label{cmetEF}
\rmd s^2= -\tilde H \rmd u^2 - 2 \rmd u \rmd r - 2A r^2 \rmd u \rmd \tilde x +\frac{r^2}{\tilde G} \rmd \tilde x{}^2 +r^2\tilde G \rmd \phi^2\ , 
\end{equation}
where
\begin{equation}
\tilde H(r,\tilde x) = 1-\frac{2M}{r}-A^2r^2 (1-\tilde x{}^2-2MA\tilde x{}^3)
                   -Ar(2\tilde x+6MA\tilde x{}^2)+6MA\tilde  x \ .
\end{equation}
The norm of the hypersurface-orthogonal Killing vector $\kappa=\partial_u$ is determined by $\tilde H$, 
\begin{equation}
\kappa_\alpha \kappa^\alpha = r^2 \tilde F=\frac{\tilde H}{ A^2}\ ,
\end{equation}
so that this Killing vector is timelike for $\tilde H>0$.

It is convenient to work with the Boyer-Lindquist-like coordinates $(u,r,\theta,\phi)$, 
obtained by defining    $\tilde x=\cos \theta$, so that
$(r,\theta , \phi)$ are spherical-like coordinates. The C metric then takes the form
\begin{equation}
\label{cmetu}
\rmd s^2= -H \rmd u^2 - 2 \rmd u \rmd r + 2A r^2 \sin \theta \rmd u \rmd \theta 
            +\frac{r^2\sin^2\theta }{G} \rmd \theta^2 +r^2 G \rmd \phi^2\ ,
\end{equation} 
where $G>0$ and $H>0$ are given by
\begin{eqnarray}
G(\theta)&=& 1-\cos^2\theta -2MA \cos^3\theta , \nonumber \\
H(r,\theta)&=&1-\frac{2M}{r}-A^2r^2 (1-\cos^2\theta-2MA\cos^3\theta) \nonumber \\
&&-2Ar\cos\theta(1+3MA\cos\theta)+6MA\cos\theta\ .
\end{eqnarray}

The C metric has event horizons (which are also Killing horizons) of the form $r=r(\theta)$ which arise as solutions of $H=0$, and these can be determined exactly.\cite{bcmprd} 
To study the location of the horizons it is useful to introduce an acceleration length scale in terms of $A>0$ given by $L_A= 1/(3\sqrt{3}A)$. It turns out that the modification of the horizons is related to the ratio
of $L_A/M = 1/(3\sqrt{3})/(AM)> 1$.
By introducing the new variable 
\beq
W= \frac{Ar}{1-Ar\cos\theta}\ , \quad
r= \frac{w}{A(1+w\cos\theta)}
\eeq
the equation $H=0$ becomes 
\beq
\label{eqW}
W^3-W+2MA=0\ .
\eeq 
For $L_A/M>1$, the solutions of this equation can be expressed as
\beq
W_1=2{\hat U}\ , \qquad 
W_2=-{\hat U}+\sqrt{3}{\hat V}\ , \qquad 
W_3=-{\hat U}-\sqrt{3}{\hat V}\ ,
\eeq
where
\beq
{\hat U}+i{\hat V} = \frac{1}{\sqrt{3}} \left(-\frac{M}{L_A} +i \sqrt{1-\frac{M^2}{L_A^2}} \right)^{1/3}\ .
\eeq
When $M\to 0$, the solution  $W_1$ gives the Rindler horizon $r=[A(1+\cos\theta)]^{-1}$, while when $A\to 0$,
the solution  $W_2$ gives the Schwarzschild horizon $r=2M$; $W_3$ leads instead to a negative value for $r$ and is therefore irrelevant.
Thus $r=[A(\cos\theta+1/W_1)]^{-1}$ and $r=[A(\cos\theta+1/W_2)]^{-1}$ correspond to the modified Rindler and Schwarzschild horizons, respectively.

\begin{figure}
\typeout{*** EPS figure 1}
\begin{center}
\includegraphics[scale=0.45]{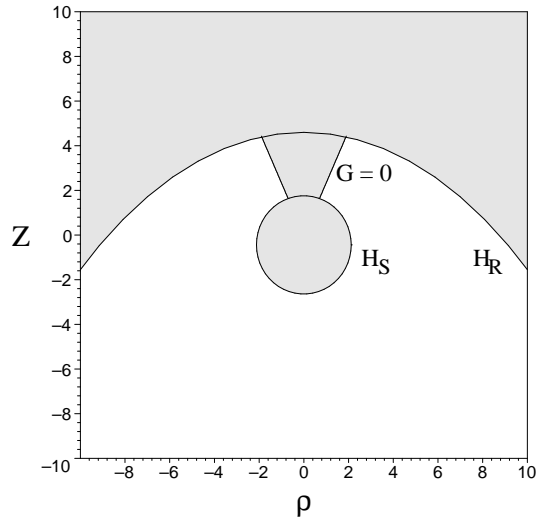}
\end{center}
\caption{ 
The accessible spacetime region (unshaded) for $MA=0.1$ is represented in the $\rho$-$z$ plane defined in terms of the polar coordinates $(r,\theta)$ as in a Euclidean plane, allowing negative values of $\rho$ to show the entire plane. The upper boundary curve $H_R$ represents the Rindler horizon while the circle $H_S$ is the Schwarzschild horizon. The forbidden conical region corresponds to negative values of the metric function $G$, i.e., to signature changes which are not considered here. 
}
\label{fig:1}
\end{figure}

The polar coordinate plane of the nonignorable coordinates $(r,\theta)$ can be represented
for visualization purposes as in flat spacetime, with horizontal and vertical axes corresponding to the usual cylindrical coordinates $\rho = r \sin\theta, z = r \cos\theta $. With this convention, the accessible region of spacetime is shown as the unshaded region in Fig.~\ref{fig:1} for $MA=0.1$.

 
In the $(u,r,\theta,\phi)$ coordinates the \lq\lq fiducial observers" are the family of static observers having 4-velocity 
$e_{\hat u}=H^{-1/2}\partial_u$ aligned with the timelike Killing vector $\partial_u$ of the boost symmetry.
A convenient orthonormal spatial frame in their local rest spaces is then
\begin{eqnarray}
\label{frame} 
e_{\hat r}=H^{-1/2}[H\partial_r-\partial_u]\ , \quad 
e_{\hat \theta}=G^{1/2}[Ar\partial_r+\frac1{r\sin\theta }\partial_{\theta}]\ , \quad 
e_{\hat \phi}=\frac{G^{-1/2}}{r}\partial_\phi\ .
\end{eqnarray}

A family of test particles moving along the $\phi$ direction with constant speed (at fixed $(r,\theta)$ are characterized by the (timelike) 4-velocity $U$ given by
\begin{equation}
\label{circolare}
U=\Gamma_{\zeta}[\partial_u + \zeta\partial_{\phi}]=\gamma [e_{\hat u} +\nu  e_{\hat\phi}]=\cosh \alpha\,  e_{\hat u} +\sinh \alpha\,  e_{\hat\phi}\ ,
\end{equation}
where $\zeta$, $\nu$ or $\alpha$ are the angular velocity, linear velocity and rapidity parametrization of this family of circular orbits respectively, related to each other by
\begin{equation}
\nu=\tanh \alpha =r(G/H)^{1/2} \zeta \ .
\end{equation}
The normalization factor $\Gamma_{\zeta}$ defined by the timelike condition $U\cdot U=-1$ is
\begin{equation}
-\Gamma_{\zeta}^{-2}
  =g_{uu}+\zeta^2g_{\phi\phi}
  = -H+\zeta^2r^2G=-\frac{H}{\gamma^2}\ ,
\end{equation}
while
$\gamma=-U\cdot e_{\hat u}=(1-\nu^2)^{-1/2}=\Gamma_\zeta H^{-1/2}$ is the Lorentz gamma factor.

The only nonvanishing components of the 4-acceleration $a(U)=\nabla_U U$ are in the 
$e_{\hat r}$-$e_{\hat\theta}$ plane (\lq\lq acceleration plane") and 
can be conveniently expressed in the form
\begin{eqnarray}\label{acccomp_gen}
a(U)^{\hat r}      =\gamma^2 k_{\rm (lie)}{}_{\hat r} (\nu^2-\nu_{(r)}^2)\ , \qquad 
a(U)^{\hat \theta }=\gamma^2 k_{\rm (lie)}{}_{\hat \theta} (\nu^2-\nu_{(\theta )}^2)\ ,
\end{eqnarray}
where
\begin{eqnarray}
\label{nurthdef}
\nu_{(r)}^2&=& \frac{M-Ar^2[GAr+\cos\theta(1+3AM\cos\theta)]}{rH}\ , \quad 
\gamma_{(r)}^2=1/(1-\nu_{(r)}^2)\ ,\nonumber \\
\nu_{(\theta )}^2&=& \frac{GAr}{GAr+\cos\theta(1+3AM\cos\theta)} , \quad 
\gamma_{(\theta)}^2=1/(1-\nu_{(\theta)}^2)\ ,
\end{eqnarray}
and the Lie relative curvature\cite{idcf1,idcf2} of the orbit 
$k_{\rm (lie)}= -\nabla\ln\,{({g_{\phi\phi}})^{1/2}}$ has been introduced, 
with frame components
\begin{equation}
\label{liecurv}
k_{\rm (lie)}{}_{\hat r} =-\frac{\sqrt{H}}{r}\ , 
\qquad  k_{\rm (lie)}{}_{\hat \theta } = -\frac{GAr+\cos\theta(1+3AM\cos\theta)}{r\sqrt{G}}\ .
\end{equation}
They satisfy the following important relations
\beq
k_{\rm (lie)}{}_{\hat r}^2 \nu_{(r)} ^2 + k_{\rm (lie)}{}_{\hat \theta }^2 \nu_{(\theta )}^2  =\frac{M}{r^3}\ , \quad 
A \sqrt{G}=-k_{\rm (lie)}{}_{\hat \theta} \nu_{(\theta)}^2\ .
\eeq
The nonnegative quantities $\nu_{(r)}$ and $\nu_{(\theta)}$ are the relative speeds at which the acceleration is either purely in the $\theta$ direction or purely radial respectively.

It is worth noting that on the surfaces where $\nu_{\rm (r)}$ diverges,  $k_{\rm (lie)}{}_{\hat r}$ vanishes in such a way that the product $k_{\rm (lie)}{}_{\hat r} \nu_{\rm (r)}^2$ is always finite there. The same is true for $\nu_{\rm (\theta )}$ and $k_{\rm (lie)}{}_{\hat \theta}$. However, while $k_{\rm (lie)}{}_{\hat r}$ vanishes on the modified Rindler and Schwarzschild horizons, there exists a surface
\beq
r =-\frac{\cos\theta(1+3AM\cos\theta)}{AG}
\eeq
(in the allowed region of the metric) where $k_{\rm (lie)}{}_{\hat \theta}$ vanishes.
It roughly corresponds to the equatorial plane of the Schwarzschild spacetime in the sense that the acceleration of circular orbits there is purely radial, although it is no longer a reflection symmetry plane as in that case.


\begin{figure}[t]
\typeout{*** EPS figure 2}
\begin{center}
\includegraphics[scale=0.45]{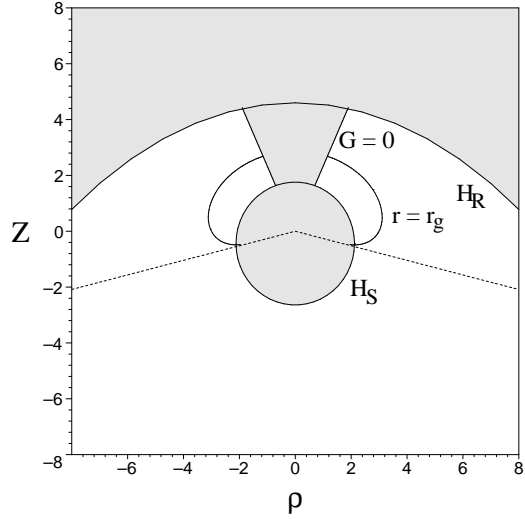}
\end{center}
\caption{The surface containing circular geodesics is shown as a curve in the $\rho$-$z$ plane indicated by the solid curve in the allowed region for $MA=0.1$.
These geodesics are timelike, null and spacelike above, on and below the equatorial plane $\theta=\pi/2$ (or $z=0$) respectively. 
The pair of dashed lines dropping down from the center correspond to the limiting value $\theta=\theta_{\rm lim}$. 
} 
\label{fig:2}
\end{figure}

\begin{figure}
\typeout{*** EPS figure 3}
\begin{center}
\includegraphics[scale=0.4]{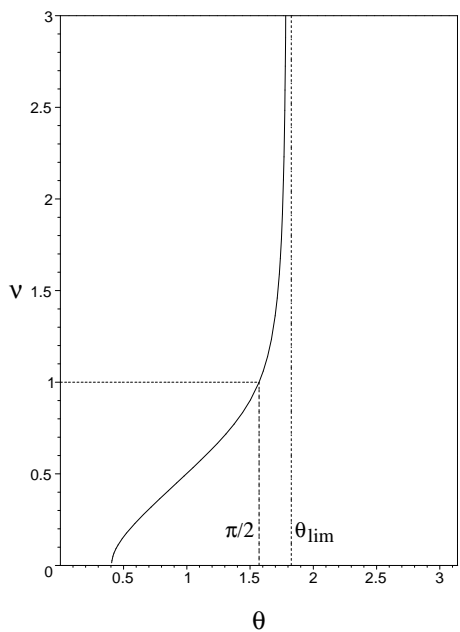}
\end{center}
\caption{The relative speed profile for the geodesics (corresponding to the positive branch $\nu_{g+}$) is plotted versus the angle $\theta$, for $MA=0.1$. The vertical dash-dotted line is at the limiting angle $\theta_{\rm lim}\approx 1.826$ (about $15^\circ$ below the equatorial plane).} 
\label{fig:3}
\end{figure}

Circular geodesics have vanishing acceleration $(a(U)^{\hat r},a(U)^{\hat \theta })=(0,0)$ when $\nu^2=\nu_{(r)}^2= \nu_{(\theta)}^2$. The latter condition determines a surface of revolution for which $r$ can be expressed as a function of $\theta$
\begin{eqnarray}
\label{rgeo}
r=r_g \equiv \frac{1}{2A}\frac{3AMG+(AM)^{1/2}[\cos\theta(4+AM\cos^3\theta)+9AM(3-2G)]^{1/2}}{G+\cos\theta[\cos\theta+3AM(2-AM\cos^3\theta)]}\ , 
\end{eqnarray}
and on this surface the relative speed can be expressed entirely as a function of $\theta$
\begin{equation}
\label{nugeo}
\nu =  \nu_{g\pm} \equiv  \pm \sqrt{\frac{Ar_gG}{Ar_gG+\cos\theta(1+3MA\cos\theta)}}\ , \qquad \gamma_{g}=1/\sqrt{1-\nu_{g\pm}^2}\ .
\end{equation} 
The surface $r=r_g$ is shown in Fig.~\ref{fig:2}. In the limit $MA\to0$ (Schwarzschild), it opens up to the equatorial plane, as shown in Fig.~\ref{fig:5} (b) (see also the discussion on the Schwarzschild limit in the next section), showing that in Schwarzschild case the timelike circular geodesics exist on the equatorial plane only  (outside the radius $r=3M$), different from the C metric case. 
Fig.~\ref{fig:3} shows, instead, the behaviour of the relative velocity (\ref{nugeo}) corresponding to the geodesics as a function of the polar angle $\theta$. Note that geodesic surface intersects the equatorial plane $\theta=\pi/2$ at $r=3M$ where $|\nu_{g\pm}|=1$, where the timelike geodesics above the equatorial plane become null.

 
The circular geodesics of the C metric, 
which represents the gravitational potential of a Schwarzschild black hole 
accelerating uniformly with acceleration $A < 1/ (3\sqrt{3} M)$ 
along the $\theta = \pi$ direction, have been obtained with the assumption that 
$A > 0$; moreover, the geodesics are timelike for $\theta_G < \theta < \pi/2$ 
and spacelike for $\pi/2 < \theta < \theta_{\rm lim}$ (see Fig.~\ref{fig:3}). 
There is a unique null geodesic for $\theta = \pi/2$ given by $r_g = 3 M$ 
irrespective of of the acceleration $A$ so long as $MA < 1/(3\sqrt{3})$. 
Here $\theta_G$ and $\theta_{\rm lim}$  depend upon $MA$, $\theta_G$ is given by $G(\theta_G)=0$ 
and $\theta_{\rm lim}$ is such that $r_g (\theta_{\rm lim})$ lies on the modified 
Schwarzschild horizon that satisfies $H ( r_g(\theta_{\rm lim}), \theta_{\rm lim} ) = 0$ (see Fig.~\ref{fig:2}). 
In the nonrelativistic limit $M\ll 1, MA \ll 1$, 
the circular geodesics reduce to the orbits that can be easily 
obtained from the Newtonian theory of gravitation: $( M/r_g^2) \cos \theta = A$ 
follows from (\ref{rgeo}) and $\nu_g^2 = (M/r_g) \sin^2 \theta$ 
from (\ref{rgeo})--(\ref{nugeo}). Furthermore, in the Newtonian limit $\zeta_g^2 = M/r_g^3$. 

In the equatorial plane there are no timelike circular geodesics. 
In fact approaching this plane ($\theta\to\pi/2$),
one has
\begin{eqnarray}
& G\to 1\ , \quad H\to 1-\frac{2M}{r}-A^2r^2\ , 
\nonumber \\
& \nu_{(r)}^2 \to \frac{M-A^2r^3}{r -2M-A^2r^3}\equiv \nu_\pm^2\ ,\quad  
\nu_{(\theta )}^2 \to 1\ , 
\nonumber \\
& k_{\rm (lie)}{}_{\hat r} \to -\frac{1}{r} \sqrt{1-\frac{2M}{r}-A^2r^2}\ , \quad 
k_{\rm (lie)}{}_{\hat \theta }\to -A \ .
\end{eqnarray}
The 4-acceleration components (\ref{acccomp_gen}) then reduce to
\begin{eqnarray}
\label{acccomp}
a(U)^{\hat r}=-\gamma^2\frac{1}{r} \sqrt{1-\frac{2M}{r}-A^2r^2}\, (\nu^2-\nu_\pm^2)\ , \quad 
a(U)^{\hat \theta }=A\ ,
\end{eqnarray}
the angular velocity associated with $\nu_\pm$ is
\beq
\zeta_\pm=\pm\left[\frac{M}{r^3}-A^2\right]^{1/2} ,
\eeq
and the relation $\zeta_\pm= - k_{\rm (lie)}{}_{\hat r}\nu_\pm $
is valid.
The velocities $\nu_\pm$, which correspond to zero radial acceleration, reduce to the geodesic velocities in the Schwarzschild limit. Instead here their acceleration is perpendicular to the equatorial plane and equal in magnitude to exactly the acceleration parameter $A$. The Lie spatial curvature vector on the other hand, as one increases the value of $A$ from 0, always has the same magnitude as in the Schwarzschild case, while its direction rotates from its initial inward radial direction downward towards $e_{\hat\theta} $. Furthermore, as one decreases the radius towards $3M$ for fixed $A$, this direction rotates downwards approaching $e_{\hat\theta} $ at the limiting maximum value $A\to1/(3\sqrt{3}M)$.

Polar coordinates $(\kappa, \chi)$ in the acceleration plane describing the magnitude of the acceleration and its orientation
\begin{equation}
a(U)^{\hat r}=\kappa \cos\chi\ , \qquad 
a(U)^{\hat \theta}=\kappa \sin \chi\ ,
\end{equation}
which are constant along any circular orbit, facilitate the transformation from the fiducial frame to the Frenet-Serret frame adapted to the orbit. Note the relation
\beq
\tan \chi = \frac{k_{\rm (lie)}{}_{\hat \theta}}{k_{\rm (lie)}{}_{\hat r}} 
             \frac{\nu^2 -\nu_{(\theta)}^2}{\nu^2 -\nu_{(r)}^2}\ .
\eeq
A Frenet-Serret frame along $U=e_0$\cite{iyer-vish} satisfying the following system of evolution equations
\beq
\label{FSeqs}
\frac{De_0}{\rmd\tau_U} =\kappa e_1\ ,\
\frac{De_1}{\rmd\tau_U} =\kappa e_0+\tau_1 e_2\ ,\
\frac{De_2}{\rmd\tau_U} =-\tau_1e_1+\tau_2e_3\ ,\ 
\frac{De_3}{\rmd\tau_U}=-\tau_2 e_2
\eeq
can then be defined by  
\begin{eqnarray}
\label{FSframe}
 e_0 &=&\cosh \alpha\, e_{\hat u}+\sinh \alpha\,  e_{\hat \phi} \ ,\
e_1=\cos \chi\,  e_{\hat r}+\sin \chi\,  e_{\hat \theta}\ ,\ 
\nonumber \\
 e_2&=&\sinh \alpha\,  e_{\hat u} +\cosh \alpha\,  e_{\hat \phi}\ ,\ 
e_3=\sin \chi\,  e_{\hat r}-\cos\chi\,  e_{\hat \theta}\ ,
\end{eqnarray}
where  $e_2=\rmd U/\rmd\alpha$, $e_3=-\rmd e_1/\rmd\chi$  and the expressions for the two torsions are given by\cite{bfdfj}
\beq
\tau_1=-\frac12 \frac{\rmd\kappa}{\rmd\alpha}
=-\frac{1}{2\gamma^2} \frac{\rmd\kappa}{\rmd\nu}\ , \qquad 
\tau_2=-\frac12 \kappa\frac{\rmd \chi}{\rmd\alpha}
=-\frac{\kappa }{2\gamma^2} \frac{\rmd\chi}{\rmd\nu}\ .
\eeq
The explicit expressions for $\kappa$, $\tau_1$ and $\tau_2$ are 
\begin{eqnarray}
\label{torsions}
\kappa &=& \gamma^2 \left[k_{\rm (lie)}{}_{\hat r}^2(\nu^2-\nu_{(r)}^2)^2
+k_{\rm (lie)}{}_{\hat \theta}^2(\nu^2-\nu_{(\theta)}^2)^2
\right]^{1/2}\ , \nonumber \\
\tau_1 &=& -\frac{\gamma^4 \nu}{\kappa} 
  \left( \frac{k_{\rm (lie)}{}_{\hat r}^2}{\gamma_{(r)}^2}
    + \frac{k_{\rm (lie)}{}_{\hat \theta}^2}{\gamma_{(\theta)}^2}\right) 
      (\nu^2-\nu_{\rm (ext)}^2)
=-\frac{\gamma^4 \nu}{\kappa} 
  \left(\kappa_{\rm (lie)}^2-\frac{M}{r^3}\right) (\nu^2-\nu_{\rm (ext)}^2)\ ,
\nonumber \\
\tau_2 &=& -\frac{\gamma^2 \nu}{\kappa} 
      k_{\rm (lie)}{}_{\hat \theta} k_{\rm (lie)}{}_{\hat r} 
          (\nu_{(r)}^2-\nu_{(\theta )}^2)\ ,
\end{eqnarray}
where
\beq
\label{nuextCm}
\nu_{\rm (ext)}^2=
\frac{(\gamma_{(\theta)}k_{\rm (lie)}{}_{\hat r} \nu_{(r)} )^2
+(\gamma_{(r)}k_{\rm (lie)}{}_{\hat \theta} \nu_{(\theta)})^2}
{(\gamma_{(\theta)}k_{\rm (lie)}{}_{\hat r})^2
+(\gamma_{(r)}k_{\rm (lie)}{}_{\hat \theta})^2}\ , \quad 
\gamma_{\rm (ext)}^2=1/(1-\nu_{\rm (ext)}^2)\ 
\eeq
is the relative velocity characterizing the extremely accelerated orbits for which $\kappa$ is an extremum among the family of circular orbits at a given location due to the vanishing of the first torsion.
The following relation holds:
\beq
\gamma_{(\theta)}^4 k_{\rm (lie)}{}_{\hat r}^2+\gamma_{(r)}^4 k_{\rm (lie)}{}_{\hat \theta}^2=\left(\kappa_{(\rm lie)}^2-\frac{M}{r^3}\right)\frac{\gamma_{(r)}^4\gamma_{(\theta)}^4}{\gamma_{(\rm ext)}^2}\ .
\eeq
By defining the orbits $U_{(r)}$ such that $\nu=\nu_{(r)}$, $U_{(\theta)}$ such that $\nu=\nu_{(\theta)}$ and $U_{\rm (ext)}$ such that $\nu=\nu_{\rm (ext)}$, the preceding relation can be written as
\beq
\frac1{\kappa(U_{\rm (ext)})^2}=\frac1{\kappa(U_{(r)})^2}+\frac1{\kappa(U_{(\theta)})^2}\ ,
\eeq 
since
\begin{eqnarray}
\kappa(U_{(r)})^2&=&\gamma_{(r)}^4k_{\rm (lie)}{}_{\hat \theta}^2(\nu_{(r)}^2-\nu_{(\theta)}^2)^2\ , \quad
\kappa(U_{(\theta)})^2=\gamma_{(\theta)}^4k_{\rm (lie)}{}_{\hat r}^2(\nu_{(r)}^2-\nu_{(\theta)}^2)^2\ , \nonumber\\
\kappa(U_{\rm (ext)})^2&=&\frac{\gamma_{(r)}^4\gamma_{(\theta)}^4k_{\rm (lie)}{}_{\hat r}^2k_{\rm (lie)}{}_{\hat \theta}^2(\nu_{(r)}^2-\nu_{(\theta)}^2)^2}{\gamma_{(\theta)}^4 k_{\rm (lie)}{}_{\hat r}^2+\gamma_{(r)}^4 k_{\rm (lie)}{}_{\hat \theta}^2}\ . 
\end{eqnarray}

Note that in the limit $\theta\to\pi/2$ one has $\nu_{(\theta)}\to\pm1$, so that $\nu_{\rm (ext)}\to\nu_{(r)}$.
On the other hand from Eq.~(\ref{torsions}) we see that $\tau_2=0$ when $\nu_{(r)}=\nu_{(\theta)}$ and this condition is satisfied on the surface $r=r_g$ defined in Eq.~(\ref{rgeo}). Therefore all the circular geodesic orbits, which lie on this surface $r=r_g$, 
have vanishing second torsion, which is the case for the equatorial plane containing the circular geodesics in the Schwarzschild spacetime.

\begin{figure} 
\typeout{*** EPS figure 4}
\begin{center}
\includegraphics[scale=0.45]{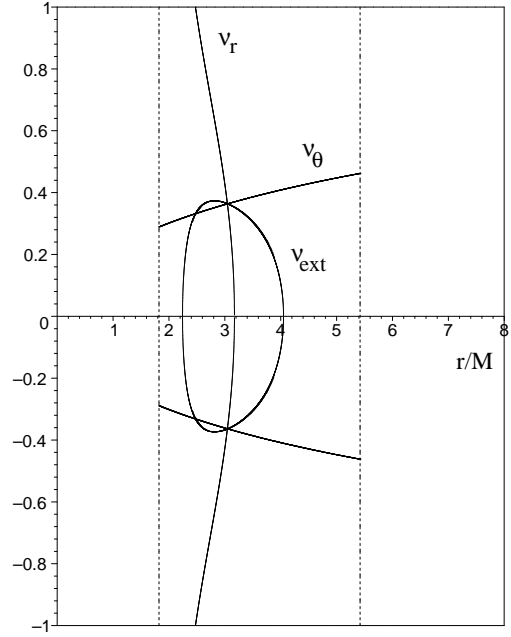}
\end{center}
\caption{The velocities $\nu_{(r)}$, $\nu_{(\theta)}$ and $\nu_{(\rm ext)}$ are plotted versus the radial parameter $r/M$ for a fixed value of the polar angle $\theta=\pi/4$, for $MA=0.1$. The vertical dash-dotted lines indicate the range of the accessible values of $r$: $1.822\le r/M\le 5.420$.
The behaviour is qualitatively the same for different values of $\theta\in[\theta_G, \pi/2)$; for $\theta=\pi/2$, instead, we have that $\nu_{(\theta)}=\pm1$ and $\nu_{\rm (ext)}=\nu_{(r)}$, while for $\theta\in(\pi/2,\pi]$ only $\nu_{(r)}$ remains timelike.
} 
\label{fig:4}
\end{figure}

Fig.~\ref{fig:4} shows the behaviours of the velocities $\nu_{(r)}$, $\nu_{(\theta)}$ and $\nu_{(\rm ext)}$ as functions of the radial distance $r$, for a fixed value of the polar angle $\theta$. Note that those extremely accelerated orbits for which $|\nu_{(r)}|=|\nu_{(\theta)}|\equiv|\nu_{g\pm}|$ are also geodesic orbits, since $\nu_{\rm (ext)}\equiv\nu_{g\pm}$ from Eq.~(\ref{nuextCm}).


On the equatorial plane, $\nu_{\rm ext\pm} = \nu_\pm$ and the Frenet-Serret scalars reduce to
\begin{eqnarray}
\kappa &=&\gamma^2 \left[\frac{r-2M-A^2r^3}{r^3}(\nu^2-\nu_\pm^2)^2+A^2(1-\nu^2)^2\right]^{1/2}\ ,\nonumber \\
\tau_1&=&-\frac{\nu\gamma^4}{\kappa}\frac{(r-3M)}{r^3} (\nu^2-\nu_\pm^2)\ , \nonumber \\
\tau_2&=&-\frac{\nu\gamma^2}{\kappa} \frac{A}{r} \sqrt{1-\frac{2M}{r}-A^2r^2}\,(r-3M)\ .
\end{eqnarray}
The timelike orbits with $\nu=\nu_\pm$ are not geodesic since $\kappa=A $ and $a(U_\pm)=Ae_{\hat \theta}$ but they are extremely accelerated, i.e., their first torsion vanishes. All the equatorial plane circular orbits instead have both vanishing first and the second torsion at the radius $r=3M$, where $\nu_\pm \to \pm 1$ and orbits with $\nu=\nu_\pm $ then correspond to null circular geodesics
and $k_{\rm (lie)}{}_{\hat r} \to -\sqrt{1/(27M^2)-A^2}$ (which vanishes at the limiting maximum value $A\to1/(3\sqrt{3}M) $) 
This special radius in the equatorial plane therefore retains the same properties that it has in the Schwarzschild spacetime as the radius at which the circular geodesics and extremely accelerated orbits merge as null geodesics with vanishing Frenet-Serret angular velocity\cite{idcf2}.

The spatial Frenet-Serret frame $e_1,e_2,e_3$ aligned with the spherical-like spatial frame rotates with respect to a Fermi-Walker transported frame along $U$ with angular velocity
\beq
\omega_{\rm (FS)}=\tau_1 e_3 + \tau_2 e_1\ ,
\eeq
the magnitude of which, $||\omega_{\rm (FS)}||=(\tau_1^2+\tau_2^2)^{1/2}$, can be directly evaluated from Eq.~(\ref{torsions})
\beq
||\omega_{\rm (FS)}||
= |\nu |\gamma^2 \left( \frac{k_{\rm (lie)}{}_{\hat r}^2}{\gamma_{(r)}^4} 
+ \frac{k_{\rm (lie)}{}_{\hat \theta}^2}{\gamma_{(\theta)}^4}\right)^{1/2}\ .
\eeq
On the equatorial plane it reduces to
\beq
||\omega_{\rm (FS)}||= |\nu |\gamma^2 (r^2-2Mr-A^2r^4)^{-1/2}|1-3M/r|\ .
\eeq
This vanishes at $r=3M$ where the acceleration of the timelike circular orbits is independent of velocity
\beq
a(U)^{\hat r} |_{r=3M} = \sqrt{\frac{1}{27M^2}-A^2}\ ,\
a(U)^{\hat\theta} |_{r=3M} = A\ ,\
\kappa |_{r=3M}=\frac{1}{3\sqrt{3}M}\ ,
\eeq
and the spherical frame vectors are Fermi-Walker transported along the orbits, as in the Schwarzschild spacetime \cite{abr}, where the null circular orbits are also geodesics at $r=3 M$. Note that the magnitude of the acceleration there has the same value as in the Schwarzschild spacetime, which coincidentally is the maximum value that the acceleration length parameter $L_A$ is allowed to take. Notice that as one increases the value of $A$ from 0, the acceleration vector rotates on a circle of this radius from its initial radial value in the Schwarzschild case to its limiting value along $e_{\hat\theta} $ when $A$ approaches its maximum allowed value of $1/(3\sqrt{3})$.

\section{The Schwarzschild and Rindler spacetime limits}

The C metric can be thought of as a nonlinear superposition of the Schwarzschild and Rindler spacetimes.
It is therefore interesting to consider both the limit $A=0$ (Schwarzschild case) and the limit $M=0$ (Rindler case) separately, showing the effects of the nonlinearity of their superposition by direct comparisons of the features of Frenet-Serret curvature and torsions, characterizing the motion of test particles along both geodesic and accelerated circular orbits. 

\subsection{The Schwarzschild limit}

The Schwarzschild metric in the $(u,r,\theta,\phi)$ coordinate system is
\begin{equation}
\label{schwu}
\rmd s^2= -\left(1-\frac{2M}{r}\right) \rmd u^2 - 2 \rmd u \rmd r +r^2\rmd \theta^2 +r^2 \sin^2\theta \rmd \phi^2\ .
\end{equation} 
The linear velocities and components of the Lie curvature simplify as follows
\begin{eqnarray}
\nu_{(r)}&=&\pm \nu_K\ , \qquad 
\nu_{(\theta)}=0\ , \qquad 
\nu_{\rm (ext)}=\pm\frac{\sin\theta}{\gamma_K}\left[\frac{M}{r-3M\sin^2\theta}\right]^{1/2} , \nonumber \\
k_{\rm (lie)}{}_{\hat r}&=&=-\frac1r \left(1-\frac{2M}{r}\right)^{1/2}=-\frac{\zeta_K}{\nu_K}, \qquad k_{\rm (lie)}{}_{\hat \theta}=-\frac1{r}\cot\theta\ ,
\end{eqnarray}
where
\beq
\label{schwgeos}
\nu_K=\left[\frac{M}{r-2M}\right]^{1/2}, \qquad 
\gamma_K=\left[\frac{r-2M}{r-3M}\right]^{1/2} , \qquad 
\zeta_K=\left(\frac{M}{r^3}\right)^{1/2} .
\eeq
In this case, no circular geodesics exist  at every radius in the equatorial plane but not at other values of $\theta$. The magnitude of the acceleration of the orbits 
reduces to
\begin{eqnarray}
\kappa &=& \gamma^2 \left[\frac{\zeta_K^2}{\nu_K^2}(\nu^2-\nu_K^2)^2 +\frac{\cot^2\theta}{r^2}\nu^4 \right]^{1/2} . 
\end{eqnarray}
All the orbits have nonzero second torsion (off the equatorial plane), while the first torsion vanishes for the extremely accelerated observers by definition
\begin{eqnarray}
\label{torsionsSchw}
\tau_1&=&-\frac{\gamma^4\nu}{\kappa}\left[\frac{\zeta_K^2}{\gamma_K^2\nu_K^2}+\frac{\cot^2\theta}{r^2}\right](\nu^2-\nu_{\rm (ext)}^2)\ , \nonumber\\
\tau_2&=&-\frac{\gamma^2\nu}{\kappa}\frac{\nu_K\zeta_K}{r}\cot\theta\ .
\end{eqnarray}

Let us now study the Schwarzschild limit of our treatment of the C metric, by expressing all relevant quantities for small values of the parameter $MA$, keeping terms only to the first order
\begin{eqnarray}
\nu_{(r)}&\simeq&\pm\nu_K \left\{1+ \frac{\cos\theta}{M\zeta_K}\left[\left(\frac{M}{r}\right)^{1/2}\frac{1}{\gamma_K^2}
 -\frac12\left(\frac{M}{r}\right)^{-1/2}\right](MA)\right\}\ ,\nonumber\\
\nu_{(\theta)}&\simeq&\pm\frac{\sin\theta}{\sqrt{\cos\theta}}\left(\frac{M}{r}\right)^{1/2}(MA)^{1/2}\ ,\nonumber\\
\nu_{\rm (ext)}&\simeq&\mp\frac{\sin\theta}{\gamma_K}\left[\frac{M}{r-3M\sin^2\theta}\right]^{1/2}
\bigg\{1+\cos\theta\bigg[\cot^2\theta-\frac{3r-(r-3M)\cos^2\theta}{r-3M\sin^2\theta}\nonumber\\
&&+\frac{r}{M^2}\nu_K^2[r-3M(1+\gamma_K^2)]\bigg](MA)\bigg\}\ ,\nonumber\\
k_{\rm (lie)}{}_{\hat r}&\simeq&-\frac{\zeta_K}{\nu_K}\left[1-\frac{r}{M}\frac{\cos\theta}{\gamma_K^2}(MA)\right]\ ,\nonumber\\
k_{\rm (lie)}{}_{\hat \theta}&\simeq&-\frac1{r}\cot\theta-\left[\frac{\sin\theta}{M}+\frac{\cot^2\theta}{r\sin\theta}(1+2\sin^2\theta)\right](MA)\ .
\end{eqnarray}

From Eq.~(\ref{acccomp_gen}), the condition for the existence of circular geodesics to first order in $MA$ becomes
\beq
0=M\cos\theta-\left[\frac{r}{\gamma_K^2\nu_K^2}+3(r-M)\cos^2\theta\right](MA)\ .
\eeq
For $MA\to0$ this condition forces the polar angle $\theta\to\pi/2$, independent of the radius $r$, so that the Schwarschild limit is exactly recovered.
As a consequence, it implies also that the surface of revolution defined by Eq.~(\ref{rgeo}), defining the locus where circular geodesics exist, must open up for decreasing values of $MA$
until it coincides with the equatorial plane in the limiting value 0.

\begin{figure} 
\typeout{*** EPS figure 5}
\begin{center}
$\begin{array}{cccc}
\includegraphics[scale=0.4]{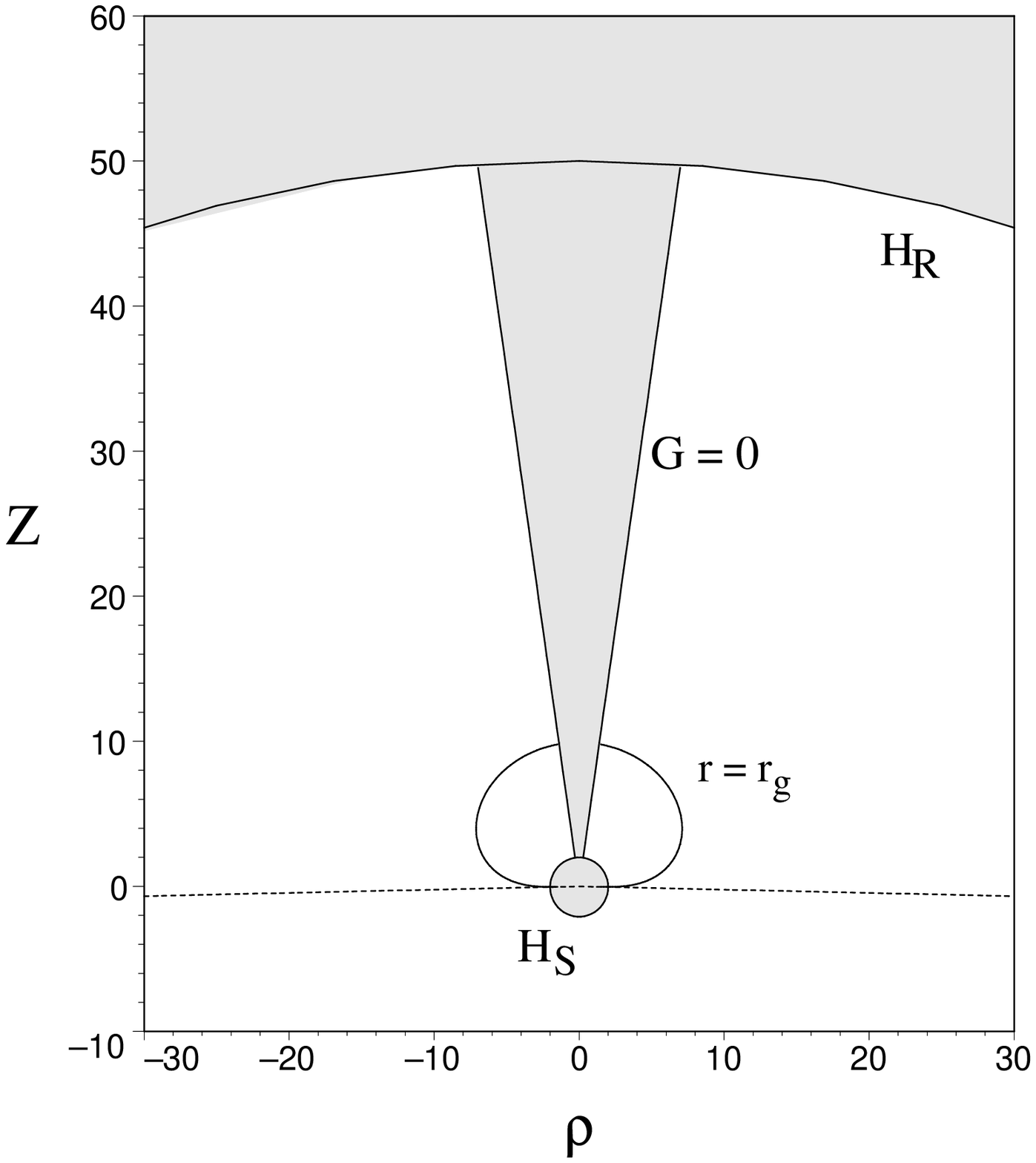}&\qquad
\includegraphics[scale=0.4]{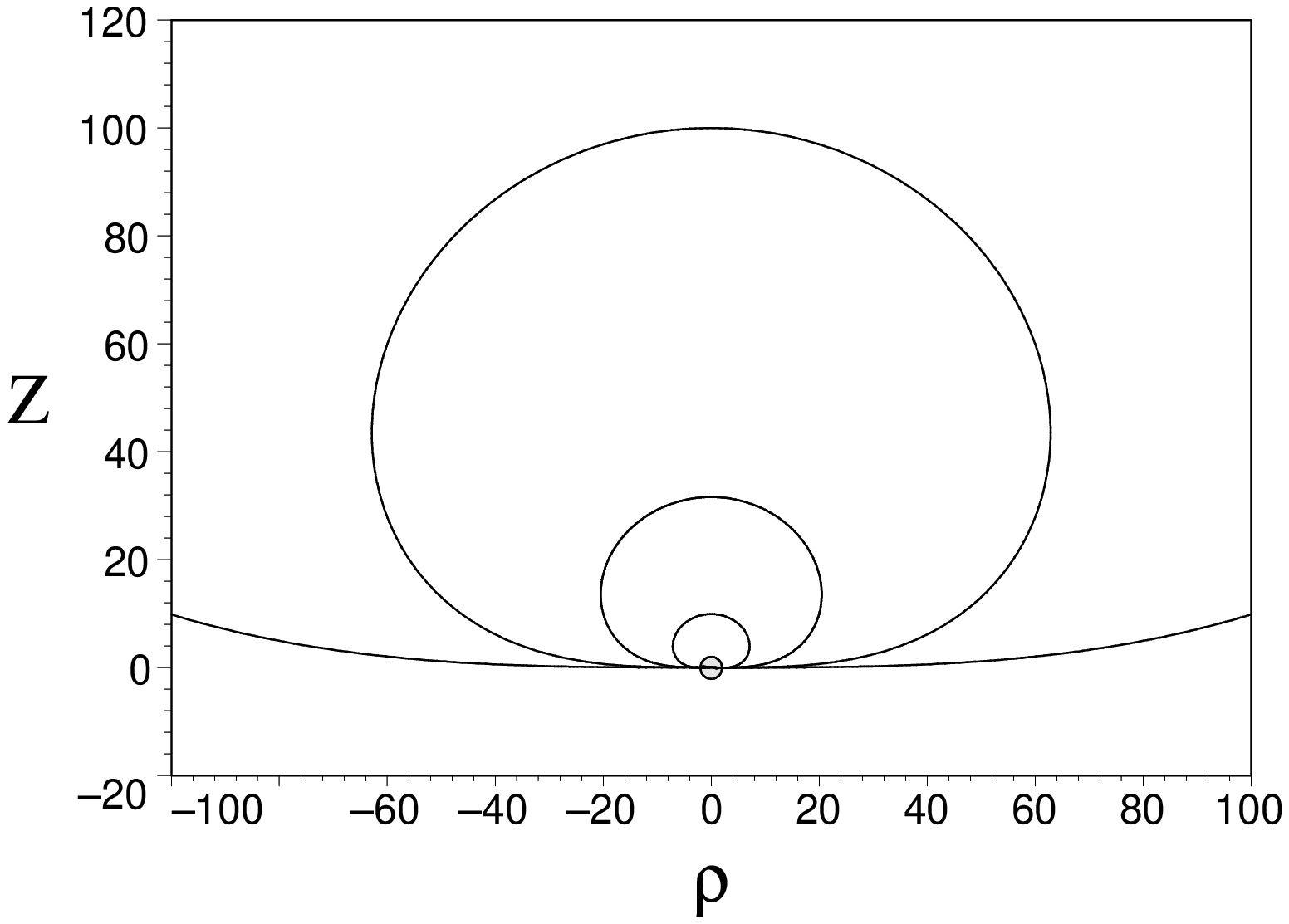}&\\[.2cm]
\qquad\mbox{(a)} &\qquad \mbox{(b)}
\end{array}
$\\
\end{center}
\caption{The surface $r=r_g$ containing the circular geodesics is plotted in Fig.~(a) for $MA=0.01$. 
The pair of straight lines dropping down from the center correspond to the limiting value $\theta=\theta_{\rm lim}$.
The same surface is plotted in Fig.~(b) for selected values  $MA=0.01, 0.001, 0.0001, 0.00001$.
The plot shows that as soon as the Schwarzschild situation is approached this surface opens up to the equatorial plane. Hence we pass from a situation in which the timelike circular geodesics exist off the equatorial plane except at one particular radius (C metric) to a situation where timelike circular geodesics exist only on the equatorial plane (Schwarzschild), at any radius $r>3M$.
}
\label{fig:5}
\end{figure}

We can get some insight into what happens when $MA\to0$ by comparing Fig.~\ref{fig:2} with the subsequent Fig.~\ref{fig:5}, where the forbidden regions as well as the surface $r=r_g$ are plotted for a smaller value $MA=0.01$. As $MA$ decreases, the modified Rindler horizon expands to infinity, the forbidden conical region shrinks away and the modified Schwarzschild horizon recovers its proper shape, while the
surface $r=r_g$ opens up to the equatorial plane, with which exactly coincides when $MA=0$.


\subsection{The Rindler limit}

The Rindler metric in the $(u,r,\theta,\phi)$ coordinate system is
\begin{eqnarray}
\label{rindu}
\rmd s^2&=& -[(1-Ar\cos\theta)^2-A^2r^2] \rmd u^2 - 2 \rmd u \rmd r + 2A r^2 \sin \theta \rmd u \rmd \theta \nonumber\\
&&+r^2\rmd \theta^2 +r^2 \sin^2\theta \rmd \phi^2\ .
\end{eqnarray} 
The horizon is located at $r=[A(1+\cos\theta)]^{-1}$.

The linear velocities and the components of the Lie curvature simplify to
\begin{eqnarray}
\label{rindkinquant}
\nu_{(r)}^2&=&-\frac{Ar(Ar\sin^2\theta+\cos\theta)}{(1-Ar\cos\theta)^2-A^2r^2}\ , 
\qquad \nu_{(\theta)}^2=\frac{Ar\sin^2\theta}{Ar\sin^2\theta+\cos\theta}\ , \nonumber \\
k_{\rm (lie)}{}_{\hat r}&=&=-\frac1r +A(Ar\sin^2\theta+2\cos\theta)\ , \qquad 
k_{\rm (lie)}{}_{\hat \theta}=-\frac1{r}\cot\theta-A\sin\theta\ .
\end{eqnarray}
The condition $\nu_{(r)}^2=\nu_{(\theta)}^2$ for the existence of timelike circular geodesics reduces to 
\beq
0=-\frac{Ar}{[(1-Ar\cos\theta)^2-A^2r^2](Ar\sin^2\theta+\cos\theta)}\ ,
\eeq
and cannot be satisfied (the surface $r=r_g$ collapses to the origin).
Eq.~(\ref{rindkinquant}) shows that the surface $k_{\rm (lie)}{}_{\hat \theta}=0$ corresponds to $\nu_{(r)}=0$.

The magnitude of the acceleration $\kappa$ for timelike circular orbits is
\beq
\kappa=\frac{\gamma^2}{r\sin\theta}\left[\nu^4+\frac{A^2r^2\sin^2\theta}{(1-Ar\cos\theta)^2-A^2r^2}\right]^{1/2}\ .
\eeq
The first and second torsions $\tau_1$ and $\tau_2$ and the magnitude of the Ferret-Serret angular velocity $||\omega_{\rm (FS)}||$ are
\begin{eqnarray}
\phantom{\omega_{\rm (FS)}}
\tau_1&=&-\frac{\gamma^4\nu}{\kappa}\frac{1}{r^2\sin^2\theta}\left[\nu^2+\frac{A^2r^2\sin^2\theta}{(1-Ar\cos\theta)^2-A^2r^2}\right]\ , \nonumber\\
\phantom{\omega_{\rm (FS)}}
\tau_2&=&-\frac{\gamma^2\nu}{\kappa}\frac{A}{r\sin\theta}\frac{1}{[(1-Ar\cos\theta)^2-A^2r^2]^{1/2}}\ , \nonumber\\
||\omega_{\rm (FS)}||&=&\frac{\gamma^2|\nu|}{r\sin\theta}\left[\frac{1-2Ar\cos\theta}{(1-Ar\cos\theta)^2-A^2r^2}\right]^{1/2} .
\end{eqnarray}
Both torsions vanish only for $\nu=0$.
In this case extremely accelerated observers do not exist.

\subsection{Conclusions}

Comparison between the features of the motion of massive particles along circular orbits in the C metric and in the limiting cases represented by the Schwarzschild ($A=0$) and Rindler ($M=0$) spacetimes shows that the situation is certainly richer in the former case. 
In fact, in the Schwarzschild case no circular geodesics can exist off the equatorial plane, since the nonradial centrifugal force cannot balance the radial gravitational attraction towards the singularity as they are not aligned. In the Rindler case circular geodesics cannot exist at all, since the centrifugal force which is never vertical cannot balance the vertical spacetime acceleration.

In the C metric the picture is more interesting. There are three forces whose balancing leads to geodesics:
1) radial attraction toward the singularity, 
2) centrifugal force, and
3) inertial acceleration along the positive $z$ direction, due to the fact that the entire spacetime is in some sense accelerating in the negative $z$ direction. 
It then turns out that geodesics can exist on a special surface $r=r_g$ (for certain values of the velocities), where the balance of the forces is realized.
This surface opens up to the equatorial plane in the Schwarzschild case (see Fig.~\ref{fig:5} (b)), while in the Rindler case it collapses to a point (the origin).
From Fig.~\ref{fig:2} as one increases the radial coordinate $r$ outside the black hole horizon, the circle of fixed $r$ intersects the surface of geodesics first at a unique value of $\theta$, and then at two values of $\theta$ until the two values coalesce and the circle passes outside this surface.

To interpret the geodesic conditions in terms  of our Newtonian perspective of relative spatial forces, it is useful to decompose the radial and angular components of the acceleration into horizontal and vertical components by introducing 
the unit vectors associated with cylindrical coordinates $(\rho,z)$ as in flat spacetime
\beq
  e_{\hat\rho} = \sin\theta\,  e_{\hat r} + \cos\theta\,  e_{\hat\theta}\ ,\  
  e_{\hat z} = \cos\theta\,  e_{\hat r} -\sin\theta\,  e_{\hat\theta} \ . 
\eeq
Then in the limit of small $M,A,\nu$ one finds the acceleration of the orbits has the form
\beq
   a_{\hat\rho} \to \frac{M}{r^2} \sin\theta - \frac{\nu^2}{r \sin\theta}\ ,\
   a_{\hat z} \to \frac{M}{r^2} \cos\theta - A\ .
\eeq
For the geodesics these go to zero, representing respectively the horizontal balance of the outward gravitational acceleration component (opposing the inward gravitational attraction towards the center) with the inward centripetal acceleration, and the vertical balance of the upward gravitational acceleration component with the downward acceleration of the reference frame of the coordinate system. Of course the exact decomposition of the acceleration relative to observers fixed in the coordinate system is complicated and not the aim of the present investigation, but it helps us bridge the strange behavior of the C metric with our Newtonian intuition.

\end{document}